On the evolution of oxidative etching of few layer graphene (FLG) in FLG /$TiO_2$ nanocomposites. Interfacial dipole signature and chemical shift in C1s X-ray photoemission spectra.

Hamza El Marouazi, Valérie Keller, Izabela Janowska*

Institut de Chimie et des Procédés pour l'Énergie, l'Environnement et la Santé, UMR 7515 CNRS/Université de Strasbourg, 25, rue Becquerel 67087 Strasbourg Cedex 2, France

Corresponding Author: *janowskai@unistra.fr, tel. +33 68852633

We present the effect of oxidative etching in FLG/$TiO_2$ hybrids on FLG edges chemistry/morphology addressing also the catalytic activity of $TiO_2$ NPs. According to the XPS and TGA analysis the oxidation degree strongly depends on $TiO_2$:FLG ratio. The preparation of two series of FLG/$TiO_2$ hybrids via sol-gel method and in-situ oxidation permitted to correctly describe and distinguish the chemical composition, core shift and charge distribution in C1s XPS with Ti2p peak as reference. Such analysis is challenging in carbon materials due the overlapping of C1s peaks from the sample and adventitious carbon reference. We claim the formation of interfacial dipole as possible direct signature of charge transfer from $TiO_2$ to FLG. A downward shift of C1s ($sp^2$C) up to 1.4 eV is measured and contributions of non $sp^2$C are determined in the composites. The modifications related to the oxidation are supported by TEM/SEM observations, and referred to non-catalytic oxidation and oxidation in mechanically mixed components.

1. Introduction

Graphene (or few layer graphene, FLG) and « nano » TiO$_2$ are both the nanomaterials of high scientific interest due to the specific properties and related potential applications. Graphene, the monoatomic sheet of hexagonal network of conjugated C=C bonds, possess high specific surface area, electrical/thermal conductivity and mechanical resistance among many others often combined properties. These allow for the potential use of graphene (or FLG) in plenty of sectors [1]. FLG being a flake containing in general less than ten graphene sheets which are stacked due to the van der Waals interactions, exhibits similar and sometimes comparable features to graphene depending on target applications. Graphene/FLG are often the additives to form different type of nanocomposites with enhanced properties, efficiency, and reactivity. Such composites are for instance graphene/FLG supported metal (oxides) that are developed and investigated in many different fields including sustainable energy development and heterogeneous catalysis [2, 3].

As to TiO$_2$, it is first of all a common semiconductor with significant photocatalytic properties in UV light range, while Ti being the second abundant metal on earth after Fe. Much less, a TiO$_2$ is known for its catalytic properties [4-6].

According to their main abilities, TiO$_2$-graphene composites are principally synthesized and investigated in view of photocatalytic applications. The first role of graphene is to accept and transport the charge formed via interactions of TiO$_2$ with photons, and consequently reduce the recombination of created electron-hole pairs [7-10]. Most of such hybrids contains graphene oxide (GO) or reduced graphene oxide (rGO) possessing highly defected carbon lattice substituted by oxygen groups. A moderate amount of such defects/functionalities is usually a benefit for dispersion and stabilization of metal (nano) particles, but if too high, the –C=C– network and related properties at longer range are disturbed. The design and control of defects/functionalities nature as well as their localization are very important in order to enhance the efficiency of graphene composites as well as the activity/reactivity of graphene it-self [11].

Despite the fact that X-ray photoelectron spectra (XPS) became a principal analysis of chemical composition of nanomaterials, a correct determination of chemical shift is not always evident [12]. This concerns especially the systems where the interfacial dipole layer is formed [13] and the chemical shifts can be easily misinterpreted and confused with electrostatic shift and vice-versa [12, 13]. In nanomaterials containing carbon of which signature overlaps with reference adventitious C this issue is more accentuated. In the case of GO (or rGO) a significant content of surface oxygen is reflected by separated peak of C bonded to the more electronegative oxygen groups (C-O-C) [14, 15] allowing for easier interpretation, which is not a case of less oxygenated carbon materials. Since the C1s position is commonly accepted as reference for determination of chemical element core position, one can find the recent report critically and precisely illustrating an abuse of such referencing [16, 17].

Herein we report the oxidative catalytic etching of FLG within the series of hybrids containing different $TiO_2$: FLG ratio. These hybrids were recently reported by us as being active photocatalysts for hydrogen production in the photoreforming of methanol. The photocatalytic activity of these hybrids depended first of all on the FLG content and related transport properties but the activity was also altered via the edges defects introduced during oxidation. This catalytic oxidative etching and its effect on the chemical composition of FLG is here investigated in details via XPS supported by TGA analysis, work function measurements and morphology observation via TEM/SEM. We claim the formation of interfacial dipole in XPS spectra that possibly confirms the charge transfer phenomena. Likewise, we distinguish the chemical shift from the electrostatic interfacial interactions in the broadened C1s spectra that correlates with oxidation degree. Up to now the $TiO_2$-hybrids with strong interfacial interactions i.e. creation of Ti–C bonds with clearly visible Ti–C peak in XPS spectra were reported but no charge transfer was observed via XPS analysis [7].

The catalytic role of $TiO_2$ NPs is clearly reflected in the activation energy values in the composites and $TiO_2$ free sample.

2. Material and methods

The synthesis of FLG, FLG', $TiO_2$ and their composites were detailed previously [10, 18, 19]. In brief, FLG and FLG' were obtained via exfoliation of expanded graphite in water in the presence of bio-surfactant (bovine serum albumin) with help of mixing-assisted sonication. FLG' is the material with higher defects content (lower size of the flakes) due to the harsher parameters of the synthesis.

$TiO_2$ and FLG/$TiO_2$, FLG'/$TiO_2$ nanocomposites were obtained via sol-gel method using precursor solution, Tetraisopropylorthotitanate ($Ti(OC_3H_7)_4$, (TIP) /acetic acid /ethanol/water, and ethanol suspension of FLG/FLG' in the case of the composites. After gelation and drying, the products were calcined at 450°C under air for 3h. FLG/$TiO_2$ contain 0.5, 1.0, 4.5 and 11.0 % of FLG (0.5%FLG/$TiO_2$, 1%FLG/$TiO_2$, 4.5%FLG/$TiO_2$, 11% FLG/$TiO_2$), while the FLG'/$TiO_2$ nanocomposites contain 0.5, 2.0 and 3.5 % of FLG' (0.5%FLG'/TiO$_2$, 2%FLG'/TiO$_2$, 3.5%FLG'/TiO$_2$).

X-ray photoelectron spectroscopy (XPS) analysis was carried out in a Thermo-VG scientific spectrometer equipped with a CLAM4 (MCD) hemispherical electron analyzer. The Al Kα line (1486.6 eV) of a dual anode X-ray source was applied as incident radiation. High resolution spectra were measured in constant pass energy mode (20 eV). The electron emission angle was 90% to the surface with 5 mm$^2$ of analyzed area. The based pressure during analysis was $5 \times 10^{-9}$ mbar and no charge neutralizer was used.

For the workfunction measurements by UPS, the HeI source at 21.23 eV was used and a bias of 15.31 V was applied to the sample, in order to avoid interference of the spectrometer threshold.

Thermogravimetric analyses (TGA) were performed using a TA Instrument Q5000IR. The samples in a platinum crucibles were heated from room temperature up to 1000 °C with a heating rate of 10 °C.min$^{-1}$ under air flow of 25 mL.min$^{-1}$.

Transmission Electron Microscopy (TEM) was performed on a JEOL 2100F LaB6 microscope operating at 200 kV and with a point-to-point resolution of 0.21 nm. Prior to the measurements, a drop of EtOH suspension of a given sample was deposited onto the gold grid covered with carbon membrane.

3. Results and Discussion

Two series of FLG/TiO$_2$ hybrids are synthesized via sol-gel method. The series II contains FLG constituted of the flakes with lower average size (FLG') as the effect of the modified conditions used during the FLG' preparation i.e. exfoliation process. It entails consequently slightly higher edges-to-surface ratio in the FLG' flakes. The FLG -TiO$_2$ precursors obtained via sol-gel step are next submitted to the calcination treatment at 450°C providing well crystalized anatase phase $TiO_2$ NPs and in-situ etched/oxidized FLG. The oxidative etching occurs in principle at the FLG edges as it is observed by TEM, SEM microscopies. According to the XPS spectroscopy and TGA analysis the degree of oxidation strongly depends on the $TiO_2$: FLG ratio. It drastically decreases with the increase of FLG amount highlighting catalytic activity of $TiO_2$. In order to confirm the catalytic role of $TiO_2$, the FLG was also oxidized alone (FLG-450°C). Likewise, the reference samples of $TiO_2$ and FLG are prepared. The related investigations of TGA results including activation energy values are presented below.

Concerning the XPS spectra, all composites demonstrate a reciprocal shift of Ti2p and C1s peak (Ti2p or C1s), fig. 1. (The general XPS spectra are in fig. S1). In this work we use only initially the C1s position at c.a. 284.8 eV as a "impurity carbon from air/adventitious" to fix the position of Ti2p in $TiO_2$ sample. Next, the same nature of $TiO_2$ NPs in $TiO_2$ sample and in the composites permitted to apply Ti2p as a reference for the composites. With this calibration a clear downward shift of the C1s peak is observed indicating the phenomena of interfacial dipole formation. (The migration of only one peak, C1s or Ti2p, excludes a considerable surface charging if any. Likewise, the unmodified Ti2p shape in the composites compared to Ti2p in $TiO_2$ indicates that no differential charging occurs, fig. S2). The interfacial dipole can be indeed formed and affected by charge redistribution factors including charge transfer, mirror force, surface rearrangement, chemical interactions, permanent dipole and interface state, which in the case of metal-inorganic semiconductor, is the metal-induced gap state (MIGS) [20, 21]. Beside XPS down-shift of C1s in the composites, we found the shift of vacuum level, i.e. change of work function (*W*) via UPS spectroscopy. The *W* of $TiO_2$ was measured to be 4.7, while it decreased to 4.3 eV for 0.5% FLG/$TiO_2$ and 1% FLG/$TiO_2$; and to 3.8 eV for 0.5% FLG'/$TiO_2$.

Two main factors are responsible for *W* decrease and this is the charge distribution at the interface and the effect of defects amount at the FLG edges. The defects (mostly FLG edges) are more abundant in series containing FLG' (0.5% FLG'/TiO$_2$). According to the work of G. Greczynski et al. the adventitious carbon layers were found to align to the vacuum level in different metals (oxides, nitrides). As a consequence, the sum of measured C 1s peak position (with respect to the spectrometer Fermi level) and the work function was constant allowing to determine correctly the position of C1s [17]. In the present work, it was estimated that the amount of adventitious C in composites is similar to the one in TiO$_2$ and this contribution was considered for deconvolution of all the spectra, fig. 1 and 2 (based on the relative amount of Cadv and Ti2p peaks in TiO$_2$, fig. S2). According to this, although approximative, quantification the general tendency of the ratio between C adv. and carbon from FLG fits well with FLG content changes.

The maximum downward shifts of C1s by 1.4 eV and 1.2 eV compared to FLG alone is measured for 11% FLG/TiO$_2$ and 3.5 % FLG'/TiO$_2$ i.e. the composites with the highest FLG content within the series (a possible C1s shift in less important manner accompanied by upwards shift of Ti2p cannot be excluded). The sp$^2$ C in more oxidized composites is less shifted due to the additional effect of conductivity loss. In 0.5% FLG/TiO$_2$ for example this "back-shift" is of 0.5 eV, what compared to FLG gives still the negative shift of 0.9 eV. The "back-shift" can be somehow explain also by intercalated oxygen as it is a case of oxygen intercalated metal-graphene interactions explained below.

Taking into account the nature of such hybrids and the literature data we suppose that observed interfacial dipole formation can be related to the charge transfer. The charge transfer phenomenon observed directly via the shift of peaks in XPS spectra was observed previously in only very few examples, and for instance in $TiO_2$-Au hybrid, where the magnitude of Au peak down-shift depended on Au thickness [22]. Other example consists of Ni-C thin alloy, where upward and downward shifts of binding energies were respectively observed for Ni 2p and C1s [23]. Finally, an upwards shift of Ti2p peak existed in the XPS spectra of $TiO_2$-GO hybrid, and its electron transfer channel character was determined using supporting analytical techniques such as X-ray absorption spectroscopy (XAS) [8]. From the other hand, the work reporting strong interfacial charge transfer (SICT) with visible Ti-C peak appearance reports no specific alternation of peaks position [7] but such shift was described in the films with co-deposited layers with no new bonding formation. In this work we can find the formation of C-O-Ti bonds, which however does not or participate in negligible manner to the interfacial dipole formation. The fact that the charge transfer in the present composites occurs is somehow confirmed earlier via enhanced photocatalytic activity and in general, it is well established in this type of hybrids [9,10].

As to the character of C1s shifts in graphene films, several works dealing with the intercalation of oxygen between graphene layer and metallic substrate are reported. It includes often the substrate of graphene growth such as Ir, Co, Cu where oxygen intercalation aims to detach the graphene from the substrate [24-27]. When graphene is electronically decoupled from metal surface by intercalation of oxygen, the down-shift of C1s is observed. This disturbed phenomenon of charge transfer from metal to graphene is sometimes defined as signature of charge transfer from oxygen covered metal to graphene [24, 25].

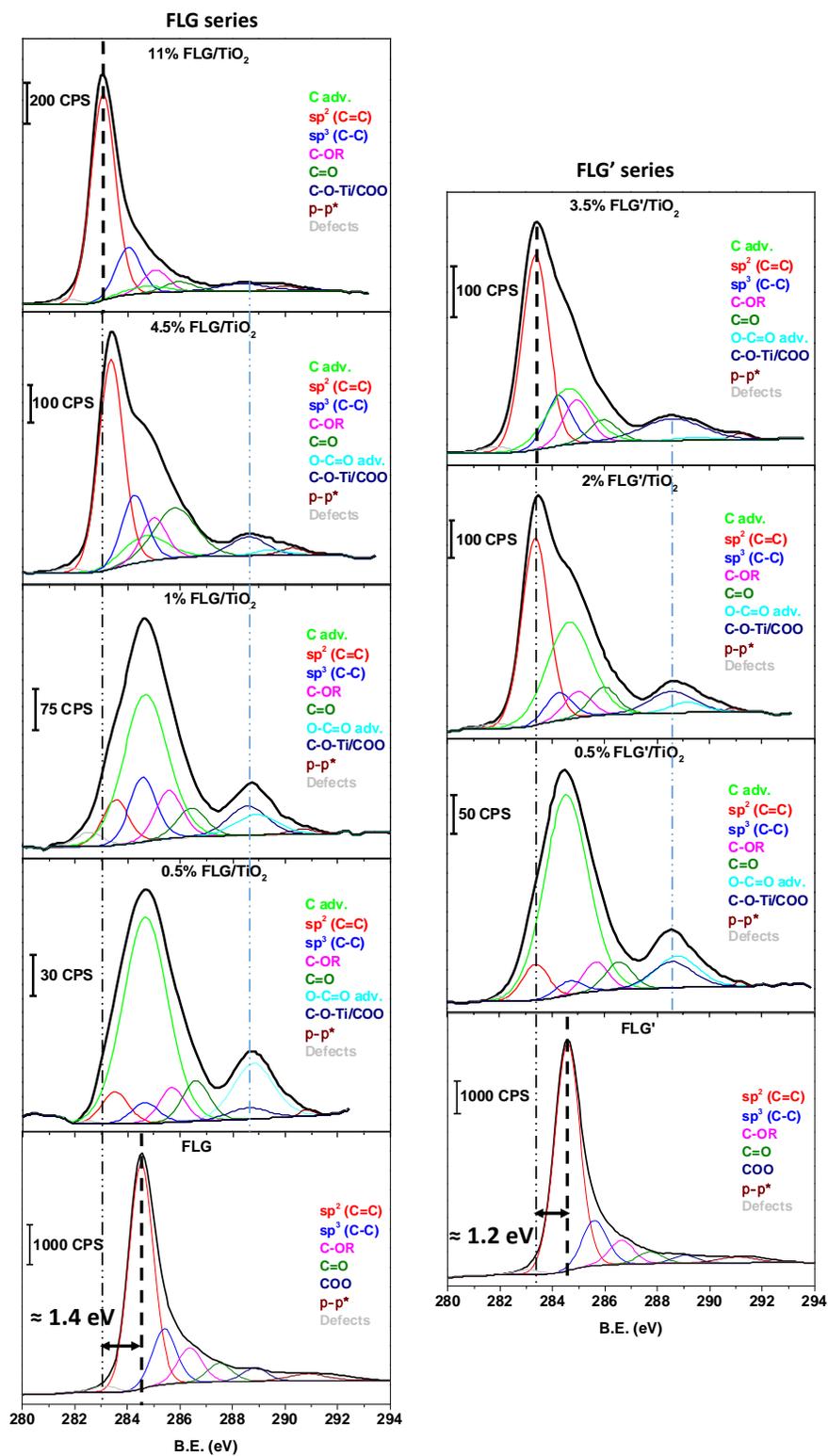

**Figure 1.** C1s XPS spectra of FLG/TiO$_2$ and FLG'/TiO$_2$ composites exhibiting downwards shift up to 1.4 eV for 11% FLG/TiO$_2$ and 1.2 eV for 3.5% FLG'/TiO$_2$ compared to FLG and FLG'.

The preparation of two series containing several composites allowed us to correctly ascribe the chemical core position of $sp^2$ and more electronegative oxygen rich C components confirming the oxidation. The broadening of C1s peak in the composites with decreasing of FLG content would be otherwise falsely interpreted as charging effect [13]. In our reasoning the FWHM of Ti 2p peak is quasi the same in all samples including $TiO_2$, while the signature of C in C1s linked to the most electronegative group such as COOH and TiOC at around 288.7 eV helps to determine the position of $sp^2$ C. This peak together with O-C=O adv.peak is progressively more pronounced with the increase of $TiO_2$/FLG ratio. Likewise, the shape of C1s peaks in the composites clearly indicates the enhancing contribution of defected C ($sp^3$C, oxygen-linked C) and C adv. In the samples with the average FLG content (4.5% FLG/$TiO_2$, 2% FLG'/$TiO_2$) the peak related to such carbons with distinguishable maximum clearly appears. These peaks decrease in 11%FLG/$TiO_2$, while it takes an advantage in the samples with low FLG content (0.5%FLG/$TiO_2$, 1%FLG/$TiO_2$, 0.5%FLG'/$TiO_2$). A detailed deconvolution of C1s peaks shows consequently that the ratio of $sp^2$C vs. the sum of defected C ($sp^3$C, C-OR, C=O, COOH) diminishes progressively in more oxidized samples. It is expected that relative TiOC interface and related oxidation degree is higher in the composites with lower FLG size since higher relative amount of $TiO_2$ is located at the FLG edges. This is reflected via higher contribution of C-O-Ti/COO peak in 0.5%FLG'/$TiO_2$ compared to 0.5%FLG/$TiO_2$. The edges are more sensitive to oxidation compared to the surface and are also favorable for the initial stabilization of metal. To check the effect of the preparation of the composites on the chemical modifications and charge redistribution induced by the catalytic oxidation with $TiO_2$, two additional samples were prepared by simple mixing of FLG' and $TiO_2$ components and next submitted to the calcination ("0.5%FLG'/TiO$_2$ mecha mix" and "3.5%FLG'/TiO$_2$ mecha mix"). The peak at 288.7 eV can be easily detected in these two samples but its intensity is weaker compared to those observed in the composites prepared *in*-situ via sol-gel method, especially due to the contribution of O-C=O adv. fig. 2. In the same context, the broadening of the main C1s is well detected in "mecha mix" samples but in less pronounced manner. The increased interface between two components in the sol-gel samples compared to the mechanically mixed find confirmation in higher ratio of C-O-Ti/COO and sp$^2$ C content.

A down-shift of C1s peak indicating interfacial dipole formation is also observed in the samples prepared by mechanical mixing. Going step further we observed no particular effect of covalent Ti-O-C bonding on down-shift of C1s peak via the sample prepared by simple brief mixing without further calcination. It confirms that the interfacial dipole formation occurs via physical metal-semiconductor contact and is not related to the formation of new covalent bonds, fig. S3.

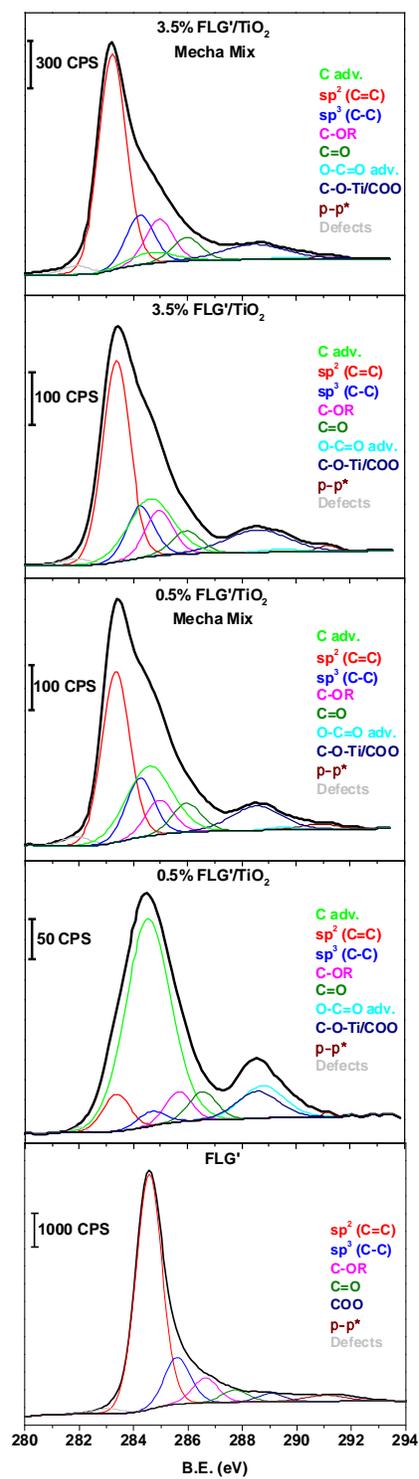

**Figure 2.** Comparative C1s spectra for 0.5 and 3.5%FLG'/TiO$_2$ composites prepared by sol-gel and mechanical mixing methods.

The strongly reduced contribution of sp$^2$C in the most oxidized composites matches with the results provided by DTG analysis, which already as such bring significant content of information (fig. 3). (The corresponding TGA curves are presented in fig. S4.) Focusing on the main combustion peak in the temperature range from c.a. 600 to 800°C, which links to the combustion of carbon, one can see that the combustion kinetics is different. The maximum combustion temperature ($T_{max}$) is shifted towards lower temperatures and the percentage of carbon burned within first few tens degrees is much higher in the most oxidized samples, demonstrating strongly diminished sp$^2$C content contrary to weakly oxidized samples, where the highest percentage of C is burned in the second half of burning region. The positive shift tendency of $T_{max}$ with increasing the sp$^2$ C is progressive, while it stabilizes and even slightly backward for the most FLG (FLG') charged samples. The latest can be related to the simultaneous phenomenon of hot spot formations during the TGA oxidation caused by the abundant FLG (FLG') well crystalized –C=C– lattice and consequently its high thermal conductivity. This was for instance observed in polymer composites containing FLG [28]. On the other hand, in the most FLG charged samples, lower relative amount of $TiO_2$ NPs will allow for easier access to carbon. As a consequence the eventual additional parasite catalytic and interfacial oxidation occurring during the analysis is in decline.

Contrary to the main carbon combustion peak, one can see the inverse tendency for the first peaks appearing below 200°C relating to the physisorption phenomena. We can ascribe the desorbing species to humidity and eventually to $CO_2$ that was produced during catalytic oxidation of FLG. The superior adsorption of such molecules on defected and oxygen-rich surfaces are indeed expected.

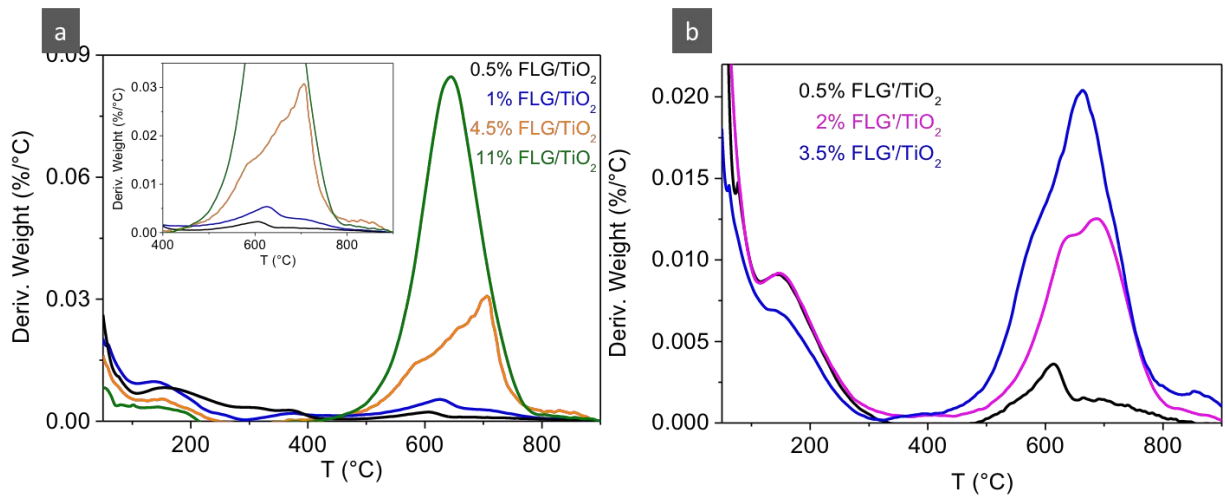

**Figure 3.** DTG curves of the composites, a) FLG-containing composites, b) FLG'-containing composites.

The quite noisy character of DTG curves comes from the fact that some additional catalytic oxidation can take place during the analysis as mentioned above. This oxidation is however insignificant taking into account a relatively fast duration of the analysis (10°C/min), and as confirmation, TGA analysis of 1%FLG/$TiO_2$ and 11%FLG/$TiO_2$ were performed additionally under inert $N_2$ atmosphere. In the DTG analysis of these two samples one can still see a significant difference in $T_{max}$ position and adequate percentage difference of carbon decomposed at lower and higher temperature range, fig. S5. Relatively smooth DTG curves of FLG, FLG' and FLG-450°C were indeed obtained as well, fig. S6. As mention above and for the reason of comparison the FLG was submitted to oxidation in the absence of $TiO_2$ (FLG-450°C). According to the TEM, TGA, fig. S7, and XPS analysis published earlier [10], the oxidation in FLG-450°C proceeds in gentler manner providing pure FLG surface with regularly etched jaggy edges. Consequently, the maximum combustion temperature via DTG for FLG 450°C is measured at almost 750°C (FLG and FLG' after synthesis are covered by bio-surfactant, that is almost removed after calcination).

Based on the TGA data, activation energy (Ea) and half-time of oxidation reaction ($t_{1/2}$) were determined for all hybrids and FLG 450°C sample, table 1. [29, 30] Ea was obtained from the slopes of Arrehenius straight plot of the ln of weight loss percentage vs. 1/T, fig. 4 a and c. In samples with low FLG content the two slopes could be distinguished with two corresponding Ea: $Ea_a$ and $Ea_b$. The second kinetic parameter, $t_{1/2}$, was obtained from the equation: $t_{1/2} = 0.693/k$, with k via the plotting the ln (1- weight loss) vs. time.

The Ea of oxidation reaction in all composites, table 1, is incomparably inferior to $E_a$ of FLG 450°C. It decreases also in significant manner along with a decrease of FLG content within the two series. In the case of the composites with highly oxidized FLG the splitting of Ea into $Ea_a$ and $Ea_b$ estimated in two regions correspond to the oxidation of highly etched carbon at the initial temperature range and oxidation of less defected carbon in the remaining temperature regime. Clearly the enhanced amount of defects facilitates the reaction. One can also see a clear impact of Ti-O-C interfacial components since Ea of the sample prepared by mechanical mixing, i.e. 3.5% FLG'/ $TiO_2$ Mecha mix is much higher compared to Ea of 3.5% FLG'/ $TiO_2$.

As to $t_{1/2}$, FLG-450°C and 3.5% FLG'/ $TiO_2$ Mecha mix are faster oxidized than the composites, which is probably linked to the easier diffusion of the gaseous reactants and thermal conductivity of FLG. In the composites, the $t_{1/2}$ increases with the increase of FLG content until certain limit achieved in highly loaded sample, 11 % FLG/$TiO_2$, where very fast oxidation can be related to the high thermal conductivity of FLG with accordance of DTG curves.

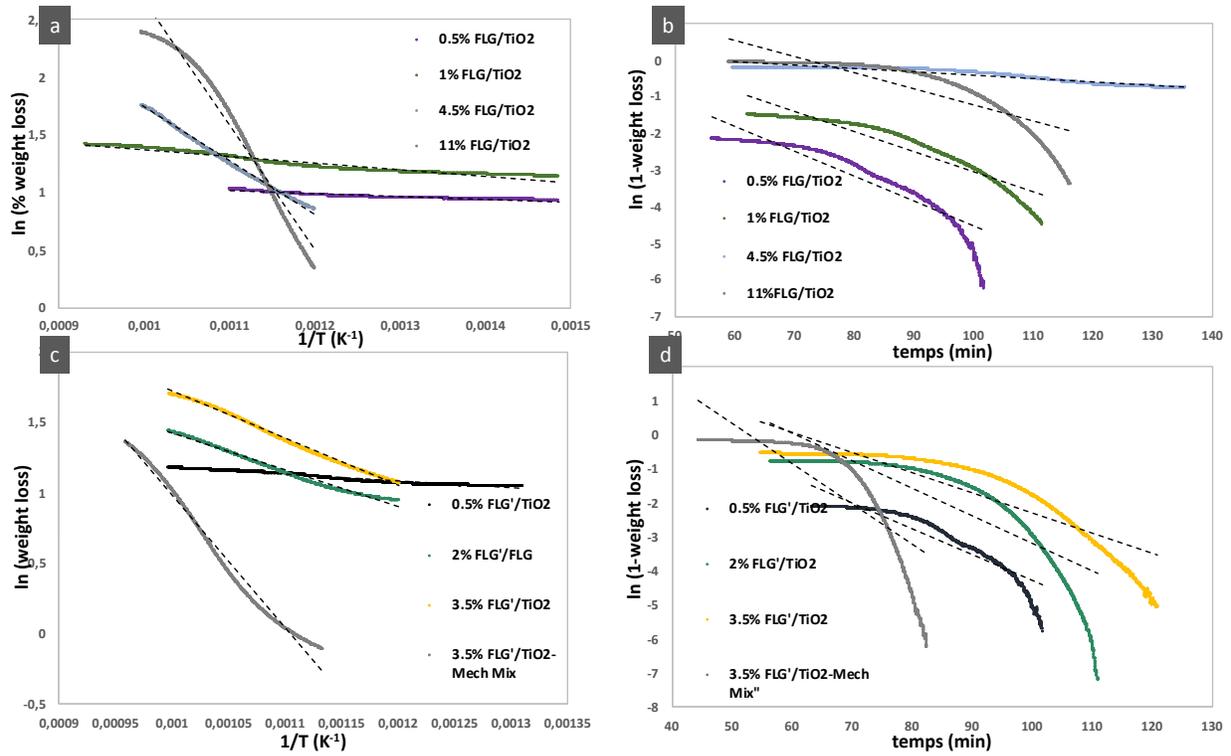

**Figure 4.** a and c) the plot of ln (%weight loss) vs. 1/T for the FLG/TiO$_2$ and FLG'/TiO$_2$ series, b and d) the plot of ln (1-weight loss) vs. time for the FLG/TiO$_2$ and FLG'/TiO$_2$ series.

Reported previously morphology studies confirm higher degree of etching in most oxidized composites occurring mainly at the FLG edges. Some TEM and SEM images are added below to support the chemical composition investigations, fig. 5.

Table 1. Kinetic parameters of oxidation reaction in FLG 450°C, FLG/TiO$_2$ and FLG'/TiO$_2$ composites

| Samples | Ea [kJ/mol] | Ea $_a$ [kJ/mol] | Ea $_b$ [kJ/mol] | t(1/2) [min] |
|---|---|---|---|---|
| *FLG 450*• | 199.28 | 265.62 | 171.65 | 4.99 |
| *0.5% FLG/TiO$_2$* | 2.00 | 1.05 | 3.94 | 10.25 |
| *1 % FLG/TiO$_2$* | 4.91 | 2.31 | 5.07 | 12.67 |
| *4.5% FLG/TiO$_2$* | 38.51 | | | 77.88 |
| *11 % FLG/TiO$_2$* | 89.69 | | | 10.65 |
| *0.5% FLG'/TiO$_2$* | 4.09 | 3.52 | 3.36 | 9.13 |
| *2 % FLG'/TiO$_2$* | 21.95 | | | 8.59 |
| *3.5% FLG/TiO$_2$* | 27.92 | | | 11.71 |
| *3.5% FLG'/ TiO$_2$ mech mix* | 78.34 | | | 5.86 |

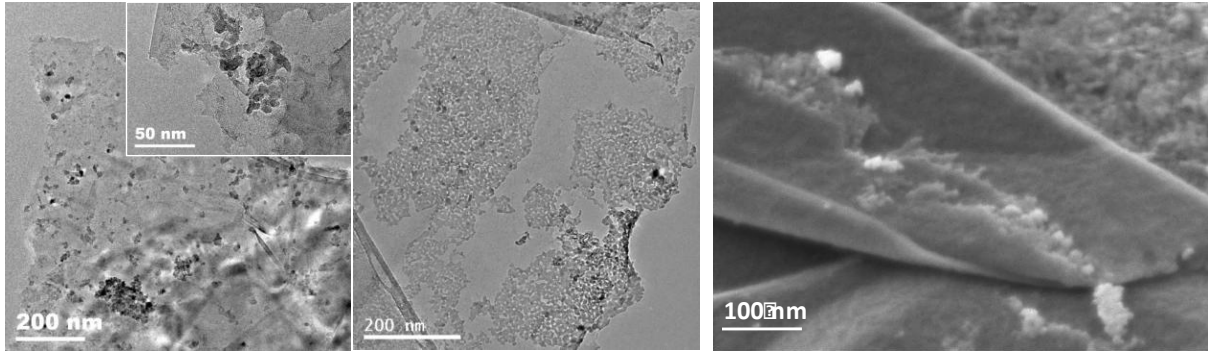

**Figure 5.** TEM (left, middle) and SEM (right) images of FLG/TiO$_2$ composites. Highly etched/oxidized irregular edges can be observed in more oxidized samples (left in-set: 0.5% FLG/TiO$_2$), middle: 11% FLG/TiO$_2$.

Additional DTG curves of the composite prepared by sol-gel method and mechanical mixing (3.5% FLG'/$TiO_2$) illustrating the enhanced C-O-Ti interface in the former is included in SI, fig. S7.

4. Conclusion

The catalytic oxidative etching of FLG in two series of FLG/$TiO_2$ composites occurring mostly at the FLG edges. Meticulous investigations of XPS and TGA demonstrated catalytic activity of $TiO_2$ and related FLG:$TiO_2$ ratio dependent degree of oxidation. The special focus on XPS spectra allowed for determination of chemical modifications, confirmation of C-O-Ti interface and the formation of interfacial dipole having possibly the origin in charge transfer from $TiO_2$ to FLG. The latter was claimed according to the downward shift of C1s vs. reference Ti 2p peak. The correct deconvolution of C1s and determination of carbon type content was achieved which is challenging in materials containing carbon. Via presented FLG oxidation we highlight additionally how different is the present oxidative etching of graphene and oxidation of graphite commonly used to obtain GO (rGO). This links primary to the localization of oxygen rich groups entailing different graphene properties.

**Supporting information** includes 7 figures: Selected general XPS spectra and XPS Ti 2p spectra, DTG curves and analysis of FLG 450°C sample.

ACKNOWLEDGMENT

Dr. Vasiliki Papaefthymiou is acknowledged for performing XPS and UPS measurements. Dr. Spiros Zafeiratos is acknowledged for UPS facility set-up. Dr. Walid Baazis is acknowledged for performing TEM measurements in IPCMS, UMR 7504 CNRS-Unistra.

Funding: This work was supported by the ANR program (PHOTER) and statutory CNRS/Unistra funding.

AUTHOR CONTRIBUTION

H. El Marouazi: synthesis and TGA analysis, data curation, formal analysis, visualization; V. Keller: supervising, funding acquisition, I. Janowska: conceptualization, investigation, methodology, supervising, funding acquisition, writing, revising.